\documentclass[letter]{aa} 

\usepackage{natbib}
\usepackage{url}
\usepackage{graphicx}
\usepackage{txfonts}
\begin{document} 

\newcommand{\cobold}{\ensuremath{\mathrm{CO}^5\mathrm{BOLD}}}
\newcommand{\linfor}{Linfor3D}
\def\teff{$T\rm_{eff }$\,}
\def\kms{$\mathrm {km~s}^{-1}$}
\def\gtsima
{\hbox{\raise0.5ex\hbox{$>\lower1.06ex\hbox{$\kern-1.07em{\sim}$}$}}}
\def\ltsima
{\hbox{\raise0.5ex\hbox{$<\lower1.06ex\hbox{$\kern-1.07em{\sim}$}$}}}

   \title{A super lithium-rich red-clump star\\
   in the open cluster Trumpler 5\thanks{Based on observations made with ESO 
   Telescopes at the La Silla Paranal Observatory
under program ID 088.D-0045(A).}}

   \author{L. \,Monaco\inst{1},
H. M. J. \,Boffin\inst{1}, 
P. Bonifacio\inst{2},
S. \,Villanova \inst{3},
G. \,Carraro\inst{1}, \\
E.~Caffau\inst{2,4},
M. Steffen   \inst{5,2},
J. A. Ahumada\inst{6},
Y. Beletsky \inst{7},
\and 
G. \,Beccari\inst{1}
          }

\institute{
European Southern Observatory, Casilla 19001, Santiago, Chile
\and
GEPI, Observatoire de Paris, CNRS, Univ. Paris Diderot, Place
Jules Janssen, 92195
Meudon, France
\and
Universidad de Concepci\'on,
Casilla 160-C, Concepci\'on, Chile
\and
Zentrum f\"ur Astronomie der Universit\"at Heidelberg, Landessternwarte, 
K\"onigstuhl 12, 69117 Heidelberg, Germany
\and
Leibniz-Institut f\"ur Astrophysik Potsdam (AIP), An der Sternwarte 16, 14482 Potsdam, Germany
\and
Observatorio Astron\'omico, Universidad Nacional de C\'ordoba, Laprida 854, 5000
C\'ordoba,  Argentina
\and
Las Campanas Observatory, Carnegie Institution of Washington, Colina el Pino, 
Casilla 601, La Serena, Chile
}
\authorrunning{Monaco et al.}
\mail{lmonaco@eso.org}

\titlerunning{A super Li-rich red-clump star in the open cluster Trumpler 5}

   \date{Received ...; Accepted...}

 
  \abstract
   {The existence of lithium-rich low-mass red giant stars still represents 
   a challenge for stellar evolution models. Stellar clusters are 
   privileged environments for this kind of investigation.}
   {To investigate the chemical abundance pattern of the old open
   cluster Trumpler\,5, we observed a sample of four red-clump stars with 
   high-resolution optical spectrographs. One of them (\#3416) reveals extremely
   strong lithium lines in its spectrum.}
   {One-dimensional, local thermodynamic equilibrium analysis was performed on
   the spectra of the observed stars. A 3D-NLTE analysis was performed to
   derive the lithium  abundance of star \#3416.}
   {Star \#3416 is super Li-rich with  A(Li)=3.75\,dex. The lack of $^6$Li
   enrichment ($^6$Li/$^7$Li$<$2\%), the low  carbon isotopic ratio
   ($^{12}$C/$^{13}$C=14$\pm$3), and the lack of evidence for  radial velocity
   variation or enhanced rotational velocity  ($v\sin i = 2.8\,$\kms) all suggest
   that lithium production has occurred in this star through the
   Cameron \& Fowler mechanism.}
   {We identified a super Li-rich core helium-burning, red-clump star in an
   open cluster. Internal production is the most likely cause of the  observed
   enrichment. Given the expected short duration of a star's Li-rich phase, 
   enrichment is likely to have occurred at the red clump or in the immediately
   preceding phases, namely during the He-flash at the tip of the red giant
   branch (RGB) or while ascending the brightest portion of the RGB.}


   \maketitle
%

\section{Introduction}

Lithium is a fragile element that is destroyed at temperatures higher than
2.5$\times$10$^6$K, which may already be reached during the contraction of a
protostellar cloud (pre-main sequence phase) that leads to a central
hydrogen-burning star. Some depletion is known to take place in the atmosphere
of stars similar to the Sun, whose lithium abundance is less than a hundredth
the level we observe in meteorites and the interstellar medium (ISM, Population
I value). In low-mass stars, additional depletion occurs once the star's
atmosphere expands as the star evolves toward the red giant phase. At this
stage, the convection in the stellar envelope brings material to the surface
from the inner parts. This material was exposed to relatively high temperatures
and is, therefore, depleted in lithium. 

Observations firmly confirm the scenario depicted above \citep[][]{gratton04}, 
and yet the existence of low-mass stars having lithium abundances in their
atmosphere that exceed the prediction of standard stellar models (lithium-rich
giants) and even the ISM level (super Li-rich giants) indicates that these stars
must synthesize lithium in their interior. Planet engulfment can be advocated to
explain lithium-rich giants, and some suggestive cases have been presented
\citep[see, e.g.,][]{adamow12}. However, accretion of a planet or a brown dwarf
in the stellar envelope would not be able to increase the surface abundance
above the Population I value. Additionally, Li-rich giants do not show enhanced
$^9$Be abundances, which would be expected if lithium enrichment is due to
planet accretion \citep[][]{melo05}. Coupled with adequately modeled mixing in
the stellar atmosphere, the \citet[][hereafter CF71]{cameron71} mechanism is
considered a viable way to produce fresh lithium in intermediate-mass asymptotic
red giants and low-mass red giants. The stellar convective envelope is first
enriched in $^3$He after the first dredge up. The latter can be burned into
$^7$Be if it is transported to regions with high enough temperature, like the
hydrogen-burning shell. $^7$Be then decays into $^7$Li, which should be brought
to the stellar surface on a fast enough timescale to escape destruction. Lithium
will then be depleted again. The lithium-rich phase would, therefore, be a
short-lived one, which would explain the low frequency ($\sim$1\%) of these
kinds of stars \citep[][]{delareza12}.

The identification of the sites responsible for this production is, however,
challenging. Early indications supported the red giant branch luminosity bump
for low-mass stars and the asymptotic giant branch clump for intermediate-mass
stars, as the preferred evolutionary stages for lithium production
\citep[][]{charbonnel00}. There are suggestions of possible clustering of
Li-rich objects among central He-burning stars \citep[][]{kumar11}, while
indications that there are lithium-rich stars all along the red giant branch
(RGB) sequence have also emerged \citep[][]{monaco11}. Most of these studies
were, however, limited by the lack of a proper evolutionary status
determination. 

For these kinds of investigations, stellar clusters are, therefore, privileged
environments, since they are single stellar populations, whose stars share a
common distance, age, and metallicity. The evolutionary stages of cluster's
stars are also readily identified. We present here our identification of a super
Li-rich, core-helium-burning red-clump star in the old, massive open cluster
Trumpler\,5 \citep[][]{kaluzny98}.


\section{Observations, data reduction, and analysis}

\begin{figure}
\includegraphics[width=1\columnwidth]{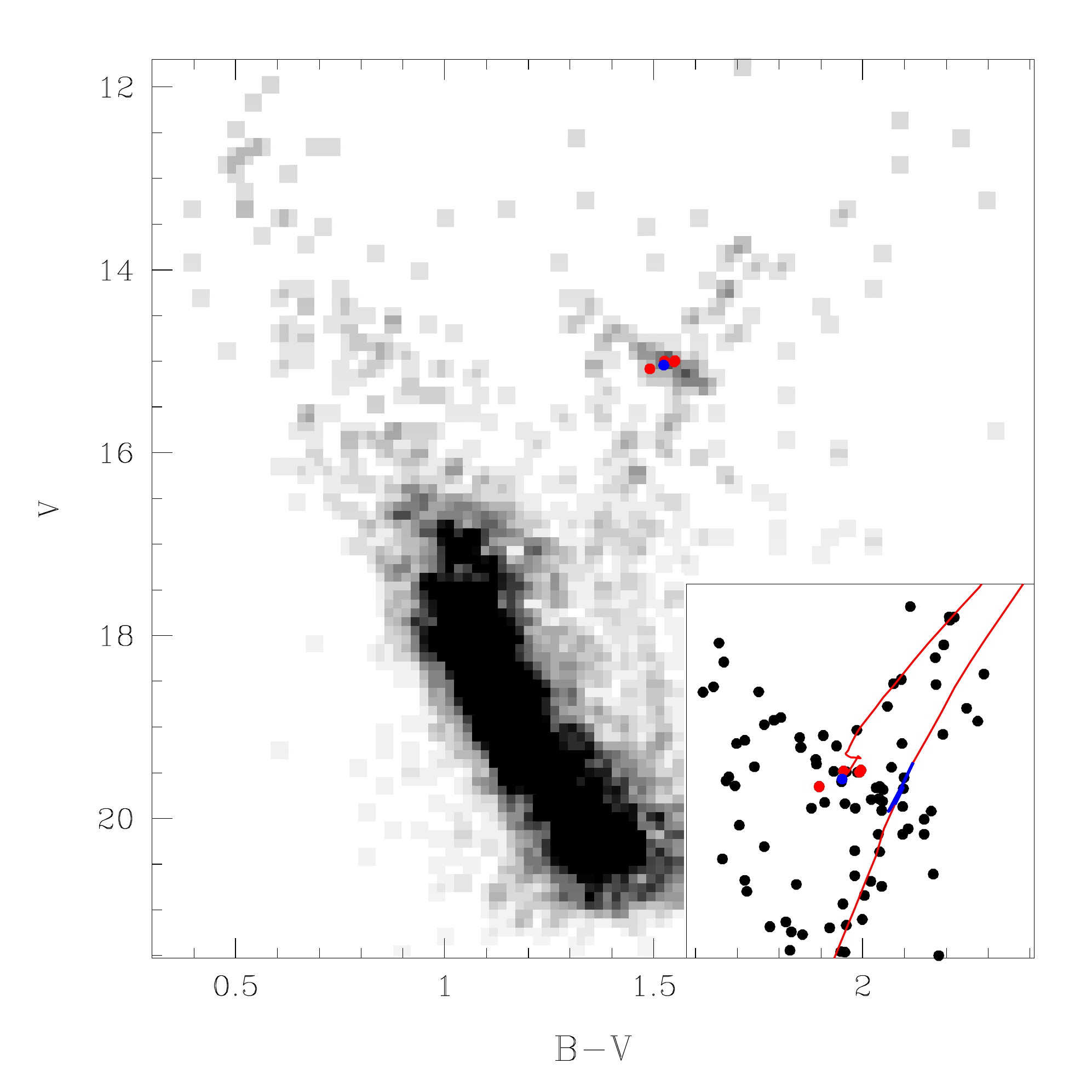}
\caption{Trumpler\,5 V {\it vs} B-V color-magnitude diagram from K98. Target
stars are marked as colored points. The super Li-rich Star \#3416 is marked in
blue. The inset presents a zoomed CMD around the clump region with an isochrone
superimposed \citep[][]{girardi02}, adopting the K98 parameters and a Z=0.008
metallicity. See Sect.\,\ref{disc} for further details.}\label{cmd}
\end{figure}

Observations of the open cluster Trumpler 5 were made during the nights of
February 11 and 29, 2012 using the multi-object fiber-fed FLAMES facility
mounted on the ESO-VLT/UT2 telescope at the Paranal observatory (Chile). Two
2400s exposures were taken simultaneously for five likely red-clump stars (see
Fig.\,\ref{cmd}) using the UVES high-resolution spectrograph red arm. The
spectra cover the 476-684 nm wavelength range and provide a resolution of
R$\simeq$47,000. Data were reduced using the ESO CPL based FLAMES-UVES
pipeline\footnote{\url{http://www.eso.org/sci/software/pipelines/}}. For Star
\#3416, additional data were obtained on October 19, 2013 with the MIKE/MAGELLAN
high-resolution spectrograph using a 0.7" slit, corresponding to a resolving
power of R$\simeq$42,000. Five 1800s exposures were gathered. The spectra were
reduced using the MIKE
pipeline\footnote{\url{http://web.mit.edu/~burles/www/MIKE/}} and cover the
wavelength range $\sim$353-940\,nm. UVES and MIKE spectra of each star were
finally corrected to the rest frame and combined. UVES and MIKE spectra of Star
\#3416 were not combined. Radial velocities were computed using the
IRAF/fxcor\footnote{IRAF is distributed by the National Optical Astronomy
Observatories, which are operated by the Association of Universities for
Research in Astronomy, Inc., under cooperative agreement with the National
Science Foundation.} task to cross correlate the observed spectra with a
synthetic one from the \citet[][]{coelho05} library with \teff=5250\,K,
log\,g=2.5, solar metallicity, and no $\alpha$-enhancement.

The local thermodynamic equilibrium (LTE) line analysis and spectrum synthesis
code  MOOG \citep[][]{sneden73} was used to perform the abundance analysis
adopting 1D ATLAS9 model atmospheres \citep[][]{k93,sbordone04}. 
Estimates of the atmospheric parameters were derived from the \citet[][hereafter
K98]{kaluzny98} photometry (see Table\,\ref{PA}). We also adopted their
reddening and distance. For the effective temperature (\teff) and
microturbulence ($\xi$), we employed the \citet[][]{alonso99} and
\citet[][]{marino08} calibrations, respectively. The stellar surface gravity
(log\,g) was derived by comparison with theoretical isochrones from the
\citet[][]{girardi02} collection with a Z=0.008 metallicity
\citep[][]{piatti04}. These initial parameters were then refined following the
procedure outlined in \citet[][]{monaco11}. Elemental  abundances were then
derived from line equivalent widths or by spectrosynthesis as required, and are
reported in Table\,\ref{Ab}  \citep[see][]{monaco11,villanova11}. Carbon,
nitrogen, and oxygen abundances as well as the $^{12}$C/$^{13}$C carbon isotopic
ratio, were determined from the MIKE spectrum for Star \#3416 alone. 

Second-epoch spectra of Stars \#4649 and \#4791 were of low quality and were,
therefore, not combined. For these two stars we analyzed spectra of only the
first epoch. Before the analysis, to increase the S/N, these two stars'
spectra were broadened adopting a Gaussian of FWHM=3.4 km/s. Given the lower
quality of their spectra, we adopted as \teff the initial value for these stars,
corrected by $-30$\,K, which is the mean difference between the initial estimate
and the values eventually adopted for the remaining stars. We classify Star
\#4649 as non-member of the cluster because its iron content differs from that
of the other stars (see Table\,\ref{PA}), and no further analysis was performed
on this star.

Lithium abundances or upper limits were derived from synthesis of the resonance
doublet at 670.78\,nm. For star \#3416, we also measured the subordinate line at
610.36\,nm. We calculated the lithium abundances corrections due to NLTE
effects, according to  \citet[][see Table\,\ref{PA}]{lind09}.

\begin{figure} 
\centering
\includegraphics[width=0.8\columnwidth]{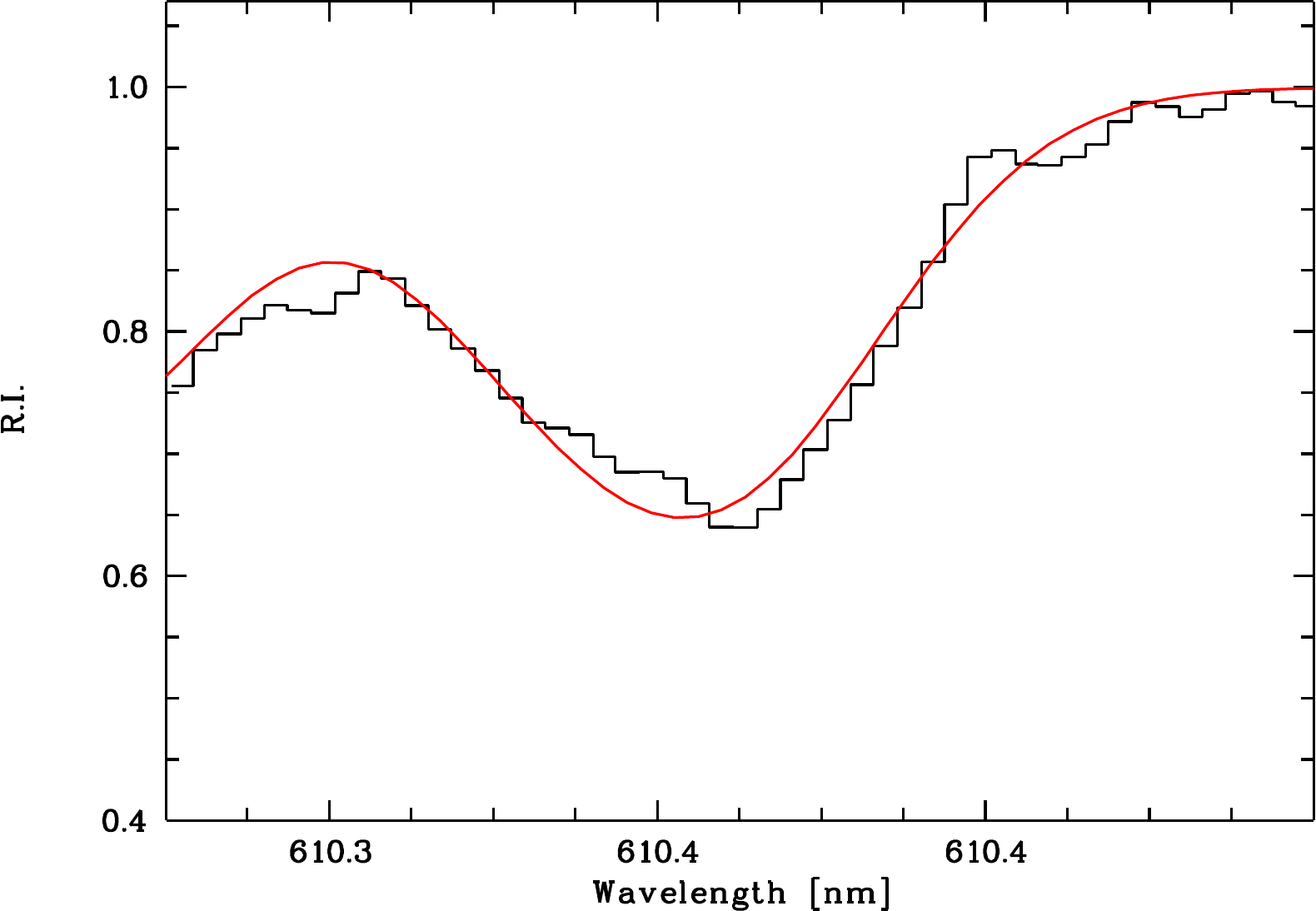}
\includegraphics[width=0.8\columnwidth]{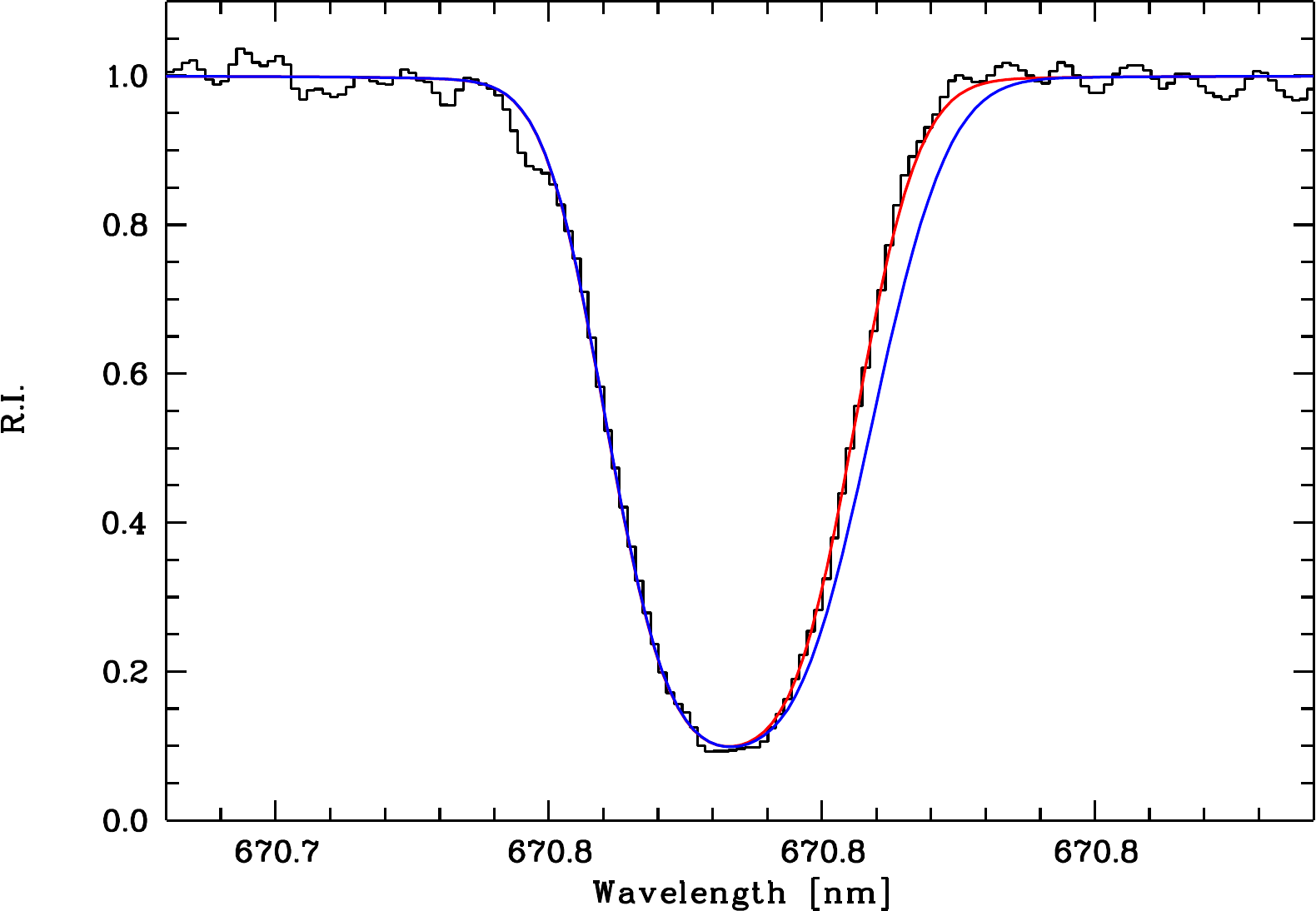}
\caption{Sample of the UVES spectrum of Star \#3416 in the  region of the Li\,I
resonance doublet (bottom panel) and subordinate line (top panel).  Synthetic
3D-NLTE spectra are shown for the same Li abundance in both panels.  In the
bottom panel also a spectrum computed with 2\% of  $^6$Li is
shown.}\label{li670b}

\end{figure}

To analyze the lithium lines in the Li-rich Star \#3416, we also made use of the
3D hydrodynamical simulations computed with the \cobold\ code  \citep[][]{FSL12}
available in the CIFIST grid \citep{Ludwig}. We computed departures from the LTE
with the {\tt NLTE3D} code of Cayrel \& Steffen,  as described in
\citet{sbordone}. The line profiles were computed with  the \linfor\ 
code\footnote{\url{http://www.aip.de/~mst/linfor3D_main.html}}. In the CIFIST
grid there is no simulation with the atmospheric parameters of Star \#3416. We
therefore used four bracketing models, with \teff=4500\,K/5000\,K and
[M/H]=0.0/$-1$, and log g = 2.5.  For each simulation we computed several
profiles with different Li abundances and then fitted the profile to the
observed UVES spectrum. In the fitting we also allowed for a Gaussian broadening
above the instrumental profile. Although the profiles from the hydrodynamical
models already contain a broadening due to macroturbulence, there may  be extra
broadening due to stellar rotation. We interpret the additional broadening as
entirely due to rotation. After deriving the Li abundance and the broadening for
each  simulation we did a bilinear interpolation to obtain the abundance and
extra broadening for the atmospheric parameters of Star \#3416. The result is
A(Li)=3.75\,dex and $v\sin i = 2.8$~\kms.  The best fit for the model with \teff
= 5000 K and solar metallicity is shown in Fig.\ref{li670b} (bottom panel). With
the broadening obtained from the resonance line we also fit the subordinate
610.3\,nm line and we obtained exactly the same Li abundance (see
Fig.\,\ref{li670b}, top panel). 

The 3D-NLTE lithium abundance derived for this star is 0.15\,dex higher than the
1D-NLTE value for the resonance 670.7\,nm doublet and 0.09\, dex higher for the
subordinate \ion{Li}{i} 610.3\,nm line.  The simplest interpretation of this
difference is that the hydrodynamical model correctly takes the microturbulence
into account, as shown by the fact that the two lines provide the same
abundance. Instead, the 1D analysis assumes too much microturbulence: a lower
assumed microturbulence would result in a higher Li abundance and a suitable
value could force the agreement between the two lines. It should  nevertheless
be noted that both lines are saturated and located on the flat part of the curve
of growth. That a microturbulent value derived from \ion{Fe}{i} lines is not
suitable for measuring lines of other elements is not surprising and simply
shows that the complexity of the velocity fields in a stellar atmosphere cannot
be modeled by a single parameter. We further notice that the 1D-LTE synthetic
profile does not reproduce the core of the 670.7\,nm line. For any interpetation
of the Li abundance in this star, the 3D-NLTE abundance should be used. 

We adopt a conservative 0.1\,dex as the abundance uncertainty corresponding to
the synthetic spectrum fitting. A $\pm$100\,K variation in the model atmosphere
\teff corresponds to a similar abundance change of about $\pm$0.1\,dex. In
Fig.\ref{li670b} we also show a synthetic spectrum computed with the same total
Li abundance,  but 2\% of $^6$Li. The extra depression on the red wing in the
latter case is obvious. We can confidently exclude the fractional ratio of
$^6$Li of this star is higher than 2\%.


\section{Discussion}\label{disc}

Stellar clusters provide easy access to a wealth of information about the basic
parameters of stars in the underlying stellar population (evolutionary stage,
age, mass, and metallicity) and are, in principle, ideal environments for 
investigating the phenomenon of lithium enrichment in giant stars.
Unfortunately, specifically devised investigations \citep[][]{pilachowski00}
have provided negative results. 

A few lithium-rich giants have been, however, identified in clusters, both open
and globular \citep[][]{carney98,hill99,kraft99,ruchti11,twarog13}.  They are
generally found along the RGB, apart for the Population\,II Cepheid V42
discovered by \citet[][]{carney98} in the globular cluster M\,5.  Here we have
presented the detection of a super-lithium-rich star (A(Li)=3.75\,dex), among
core He-burning red-clump stars in the open cluster Trumpler\,5.

Traditionally, three possible causes of lithium enrichment beyond the level
allowed by standard stellar evolution theory models have been suggested:
preservation of the original lithium content, pollution by an external source,
for instance, the engulfment of a planet or a brown dwarf and, finally, internal
production through the CF71 mechanism.

In the present case, the first possibility is excluded by the abundance higher
than the ISM level in Star \#3416. Additionally, Trumpler 5 is very similar to
the open cluster NGC2243 in terms of age and metallicity. Unevolved main
sequence stars in this cluster have lithium abundances well below the
Population\,I value \citep[A(Li)=2.70$\pm$0.20,][]{francois13}, in agreement
with the abundance measured at similar metallicity in the Small Magellanic Cloud
ISM \citep[][]{howk12}. It is, therefore, reasonable to think that Trumpler\,5
has a similar original lithium abundance. Furthermore, the low measured
$^{12}$C/$^{13}$C ratio (14$\pm$3) indicates that this star underwent the
internal mixing episodes associated with the first dredge-up and the RGB-bump
level \citep[see][]{gratton04}. This is common for Li-rich giants, which do
present evidence of the mixing they have experienced \citep[][]{charbonnel00}.

In a similar vein, lithium enrichment due to a planet engulfment episode would
not be able to raise the star's atmospheric abundance above the ISM value. This
should also be recognizable by the presence of a large fraction of $^6$Li, which
we rule out for Star \#3416. Increase in $^6$Li is also foreseen for lithium
enrichment due to stellar flares \citep{tati}, which can also be ruled out as a
cause of enrichment. Additionally, Star \#3416 presents a slow rotational
velocity (2.8\,\kms), and we did not find any indication of binarity -- the
radial velocities measured on the two UVES epochs and the MIKE observations are
compatible with no variation ($\Delta v_{\rm helio}<$1\,\kms), given the errors
in the measurements.  

A significant lithium production must then have occurred in the stellar interior
of Star \#3416. For the three other stars in Trumpler 5, we measure lithium
abundances A(Li)$<$1.3\,dex (see Table\,\ref{PA}).  Star \#3416 thus shows a
factor $\sim$280 increase in its surface lithium abundance, compared to other
red-clump stars in Trumpler\,5.

At which evolutionary phase was the lithium produced? Given the expected short
lifetime of the Li-rich phase of a star and its high abundance, we may consider
that enrichment in \#3416 appeared either during the current red-clump phase or
just before, i.e. at the tip of the RGB and its connected core He-flash or on
the brightest portion of the RGB. Trumpler\,5 red-clump stars have, in fact, a
current mass range of $\sim$1--1.4\,M$_\odot$ (see Fig.\ref{cmd}), as derived
by  the comparison with isochrones from the \citet[][]{girardi02} collection
and adopting the K98 parameters in terms of distance, reddening and age (4
Gyr). Other cases of lithium-rich giants associated with the red-clump phase
have been suggested \citep[][]{ruchti11,kumar11,adamow12b,martell13}, but,
being field stars, they lacked any firm determination of their evolutionary
status.

A preferred association of low-mass lithium-rich giants with the RGB-bump was
proposed by \citet[][]{charbonnel00}. They argue that removing the mean
molecular weight discontinuity left over by the first dredge-up occurring at
the RGB-bump led to the onset of the extra mixing between the hydrogen-burning
shell (HBS) and the envelope, allowing the CF71 mechanism to become an
effective route for producing lithium-rich giants. Indeed, it is now accepted
that the onset of thermohaline mixing at the RGB-bump will induce extra mixing
between the convective envelope and the stellar interior \citep[see][and
references therein]{eggleton08,charbonnel10}. The more robust sample studied by
\citet[][hereafter KRL11]{kumar11}, on the other hand, supports an association
with the red-clump region and lithium production associated with the He-flash
at the RGB tip. 

We notice that a number of lithium-rich and super lithium-rich giants have been
identified in the brightest portions of the RGB
\citep[][]{kraft99,monaco08,monaco11,ruchti11,martell13}, therefore the
\citet[][]{charbonnel00} or the KRL11 proposals are unable to embrace the whole
complex phenomenology of lithium-rich giants, nor do they try to account for it
\citep[see][]{charbonnel00}. Nevertheless, the existence of such a preferred
association would be an important feature. That extensive surveys for Li-rich
giants \citep[][]{monaco11,ruchti11,martell13} have failed to detect it calls
for an explanation. This is, however, beyond the scope of the present paper.
Suffice it to say here that the different studies may be covering different
parameter spaces. As an example, the bulk of lithium-rich giants in the KRL11
work have \teff between $\sim$4500 and 5000\,K and log\,g between 1.3 and 2 (see
their Fig.2, bottom panel), which is a region poorly sampled by the
\citet[][]{martell13} survey (see their Fig.\,1).

It is unclear at present if current stellar models are capable of producing a
significant amount of lithium during the He-flash. We remark here that our stars
very likely have lower masses than the KRL11 ones, and a lower amount of $^3$He might
remain in the stellar envelope to allow for the CF71 mechanism to operate, after
the depletion operated by thermohaline mixing \citep[][]{eggleton08}. Therefore,
different mechanisms may have operated to produce the lithium abundances
observed in KRL11 stars and in Star \#3416 in Trumpler\,5. 

Indeed, the \citet[][hereafter D12]{denissenkov12} models do not allow for
lithium production during the He-flash \citep[see their discussion but
also][for a possible alternative scenario]{mocak11a,mocak11b}. D12 proposes,
alternatively, that  lithium-rich stars identified as belonging to the
red clump in the KRL11 sample may, in fact, be close to the RGB bump. but that
some internal extra-mixing, perhaps related to fast internal rotation, may
cause the stars to perform an excursion in the HR diagram towards redder and
fainter magnitudes, compatible with  the expected location of red-clump stars
with masses of $\sim$2\,M$_\odot$. 

This scenario may apply to all the lithium-rich giants tentatively identified
in the field as red-clump stars. It fails, however, to explain the Trumpler\,5
case. Stars compatible with the D12 hypothesis would be found at redder colors
than the red clump and at fainter magnitudes, but this location is incompatible
with Star \#3416. While the present paper was in the refereeing stage, the
detection in the {\it Kepler} field of a core-helium-burning lithium-rich giant
(A(Li)$_{NLTE}$=2.71) confirmed by asteroseismic analysis was reported
\citep[][]{silva14}. As for Star \#3416, this giant is not carbon rich, and it
presents a low carbon isotopic ratio ($^{12}$C/$^{13}$C$<$20). These two stars,
then provide definitive confirmations of the existence of low-mass lithium-rich
and super-lithium-rich stars among red-clump stars.

The D12 model may, nevertheless, be at work in other cases. In this respect,  it
is important to notice the detection of a lithium-rich giant at a level below
the RGB-bump in the open cluster NGC6819. This location may indeed be explained
by the D12 model \citep[][]{twarog13}.  


\begin{acknowledgements}

P.B. acknowledges support from the CNRS-INSU PNCG. 
S.V. gratefully acknowledges the support provided by FONDECYT N. 1130721.
E.C. is grateful to the MERAC foundation for funding her fellowship. 
J.A.A. is grateful to ESO for supporting his visit to the Santiago
premises in 2010 and 2011, where this project started.

\end{acknowledgements}


\Online
\appendix

\section{Detailed species abundances}

Table\,\ref{PA} reports program stars' basic parameters: coordinates, magnitudes,
atmospheric parameters, radial velocities, and spectral signal-to-noise ratio.
The derived iron and lithium abundances are also reported.

Table\,\ref{Ab} reports the detailed species abundances measured for target
stars from 1D-LTE analysis. The adopted solar references are also
reported. 

\begin{table*}
\caption{Program star coordinates, photometry (from K98), adopted atmospheric parameters, and iron and lithium abundances. The spectra signal-to-noise
ratios are also indicated, as well as the measured radial velocities. Lithium abundances are derived from the 670.78\,nm line for all stars but for 
\#3416, for which the values from both the 670.78\,nm and the 610.36\,nm lines are reported as (A(Li)$_{670.78\rm \,nm}$/A(Li)$_{610.36\rm \,nm}$).}
\label{PA}
\begin{center}
\begin{tabular}{lccccccccccrl}
\hline
ID  & $\alpha$(J2000) & $\delta$(J2000)&    V	 &    B-V & \teff         &log g & $\xi$  & [Fe/H] &  A(Li)  & S/N  & $v_{\rm helio}$  \\
   &&&&  & K             &      & \kms    &     &   &  @607\,nm &\kms\\
\hline
\object{1318}	   & 06:36:53.5  & 09:25:34.6	& 15.01  &  1.55     &4730  &2.20  &1.20 &$-0.49$ & 1.23$^a$       	&  33  &  48.1$\pm$0.3$^e$      \\
\object{3416$^c$}  & 06:36:40.2  & 09:29:47.8	& 15.04  &  1.52     &4850  &2.20  &1.28 &$-0.51$ & (3.75/3.75)$^b$  	&  36  &  49.8$\pm$0.1$^e$      \\ 
                   &             &        	&        &           &      &      &     &      & (3.60/3.66)$^a$  	&      &                        \\ 
\object{3416$^d$}  & 06:36:40.2  & 09:29:47.8	& 15.04  &  1.52     &4870  &2.05  &1.33 &$-0.53$ & ---	     		& 103  &  50.5$\pm$0.2$^f$     \\ 
\object{4649}	   & 06:36:48.0  & 09:32:33.5	& 15.00  &  1.55     &4750  &2.25  &1.02 &$-0.27$ & ---	     		&  14  &  47.3$\pm$0.1$^e$ \\ 
\object{4791}	   & 06:36:33.1  & 09:33:03.7	& 15.00  &  1.53     &4800  &1.90  &1.30 &$-0.48$ & $<$1.30$^a$    	&  11  &  50.0$\pm$0.1$^e$ \\
\object{6223}	   & 06:36:38.6  & 09:38:52.6	& 15.08  &  1.49     &4830  &2.40  &1.21 &$-0.42$ & $<$1.18$^a$    	&  30  &  50.8$\pm$0.4$^e$ \\
\hline
\end{tabular}
\end{center}
$^a$ Abundance derived with 1D-LTE analysis plus NLTE corrections calculated according to \citet[][]{lind09}. Owing to the
grid boundaries, NLTE corrections were adopted assuming $\xi$=2.0~\kms. \\
$^b$ Value measured from 3D-NLTE analysis.\\
$^c$ Analysis based on UVES spectrum. \\
$^d$ Analysis based on MIKE spectrum. \\
$^e$ Radial velocity uncertainty calculated as the difference between the two epochs $v_{\rm helio}$
multiplied by 0.63 \citep[small sample statistics, see][]{keeping62}.\\
$^f$ Formal error on radial velocity determination from cross-correlation analysis. 

\end{table*}

\begin{table*}

\caption{Measured abundances for the program stars. Adopted absolute (A(X)) solar values are also reported. Number in parenthesis indicate the number of lines from which  the abundance is
derived and the abundance Gaussian $\sigma$ value, when the abundance is derived
from more then one line. The two values for the carbon and nitrogen abundances  for Star \#3416 are derived
from the G-band and C2 band at 563.5\,nm and from the CN lines at 634\,nm and 800\,nm. $^{12}$C/$^{13}$C isotopic ratio was measured from CN lines at 800\,nm.}\label{Ab}

\begin{center}
\begin{tabular}{l|l|l|l|l|l|c}
\hline
           Element               &  \#1318            &       \#3416$^a$    &        \#3416$^b$        &        \#4791           &     \#6223	 &	  Sun  \\
\hline 
         {[}Fe\,I/H]		 & $-0.49$ (107/0.12) &   $-0.51$ (95/0.11) & $-0.53$	    (112/0.07) &  $-0.48$    (33/0.14)   &     $-0.42$ (68/0.08) &	7.50   \\      
         {[}Fe\,II/H]		 & $-0.50$ (11/0.13)  &   $-0.53$ (10/0.17) & $-0.54$	     (12/0.10) &  $-0.49$    (7/0.17)	 &     $-0.44$ (9/0.09)  &	7.50   \\      
         {[}C/Fe]		 &   ---     	      &      ---     	    & $-0.17$/$-0.07$	       &    --- 		 &       ---	         &      8.49   \\  
         {[}N/Fe]		 &   ---     	      &      ---     	    &	0.05/0.15              &    ---     		 &       ---     	 &	7.95   \\  
         {[}O/Fe]		 &   ---     	      &      ---     	    & $-0.07$	    (1/--- )   &    ---     		 &       ---     	 &	8.80   \\  
         $^{12}$C/$^{13}$C	 &   ---     	      &      ---     	    &  14$\pm$3                &    ---     		 &       ---     	 &		\\  
         {[}Na/Fe]	         &   0.09 (2/0.06)    &     0.12 (1/--- )   &	0.18	    (2/0.01)   &    0.24    (1/--- )	 &       0.16 (2/0.12)   &	6.32   \\     
         {[}Mg/Fe]	         &   0.17 (1/--- )    &     0.25 (1/--- )   &	0.27	    (1/--- )   &    0.08    (1/--- )	 &	 0.17 (1/--- )   &	7.56   \\     
         {[}Al/Fe]	         &   0.26 (2/0.04)    &     0.22 (1/--- )   &	0.22	    (1/--- )   &    0.35    (2/0.11)	 &	 0.25 (1/--- )   &	6.43   \\     
         {[}Si/Fe]	         &   0.03 (6/0.18)    &     0.06 (5/0.12)   &	0.08	    (8/0.07)   &    ---     		 &	 0.03 (5/0.06)   &	7.61   \\     
         {[}Ca/Fe]		 &   0.14 (6/0.08)    &     0.16 (6/0.08)   &	0.18	    (6/0.10)   &  $-0.06$    (5/0.43)	 &       0.10 (7/0.14)   &	6.39   \\    
         {[}Ti/Fe]		 &   0.17 (8/0.11)    &     0.10 (10/0.13)  &	0.12	    (10/0.09)  &    0.16    (4/0.25)	 &       0.30 (8/0.09)   &	4.94   \\   
         {[}Ti\,II/Fe\,II]	 &   0.09 (2/0.03)    &     0.18 (2/0.22)   &	 ---	     	       &     ---    		 &	 0.14 (2/0.02)   &	4.94   \\
         {[}V/Fe]		 &   0.30 (1/--- )    &     0.32 (1/--- )   &	 ---	     	       &     ---    		 &	 0.24 (1/--- )   &	4.00   \\
         {[}Cr/Fe]	      	 &   0.00 (6/0.15)    &     0.07 (6/0.17)   &   0.16	     (6/0.09)  &     ---    		 &       0.30 (7/0.29)   &	5.63   \\   
         {[}Ni/Fe]	      	 &   0.05 (15/0.17)   &     0.02 (16/0.17)  &	0           (24/0.09)  &    0.05    (6/0.19)	 &       0.14 (18/0.19)  &	6.26   \\   
         {[}Cu/Fe]	      	 &   0.21 (1/--- )    &     0.31 (1/--- )   &   0.25         (1/--- )  &    ---     		 &       ---	 	 &	4.19   \\  
         {[}Zn/Fe]	      	 &   0.01 (1/--- )    &     ---             &    ---                   &    ---     		 &     $-0.18$ (1/--- )  &      6.61   \\
         {[}Y\,II/Fe\,II]	 & $-0.07$ (3/0.03)   &     0.04 (3/0.08)   &	   0	     (2/0.00)  &    ---     		 &     $-0.01$ (3/0.33)  &      2.25   \\ 
         {[}Zr\,II/Fe\,II]	 &   0.09 (1/--- )    &     0.08 (1/--- )   &	0.05	     (1/--- )  &    ---     		 &	 0.04 (1/--- )	 &	2.56   \\  
         {[}Ba\,II/Fe\,II]	 &   0.06 (1/--- )    &     0.08 (1/--- )   &	0.19	     (1/--- )  &    0.17    (1/--- )	 &	 0.16 (1/--- )	 &	2.34   \\  
         {[}La\,II/Fe\,II]	 &   0.09 (1/--- )    &     0.19 (1/--- )   &	 ---	               &    ---     		 &     $-0.10$ (1/--- )  &      1.26   \\
         {[}Eu\,II/Fe\,II]	 &   0.27 (1/--- )    &     0.21 (1/--- )   &	0.18	     (1/--- )  &    0.19    (1/--- )	 &       0.07 (1/--- )   &      0.52   \\   
\hline
\end{tabular}
\end{center}
$^a$ Analysis based on UVES spectrum. \\
$^b$ Analysis based on MIKE spectrum. \\
\end{table*}

\end{document}